\documentclass{mn2e}
\usepackage{psfig}
\def    \kms    {\rm km\ s^{-1}} 

\def    \modot  {\rm M_{\odot}}

\begin{document}

\title[Massive WR binary SMC WR7]
{The Massive Wolf-Rayet Binary SMC WR7 }

\author[V.S. Niemela et al.]{ V.S. Niemela $^1$ 
\thanks {Member of Carrera del Investigador, CIC--BA, Argentina;
Visiting Astronomer, CTIO, NOAO, operated by AURA, Inc., for NSF;
Visiting Astronomer, CASLEO, San Juan, Argentina},
P. Massey $^2$
\thanks {Visiting Astronomer, CTIO, NOAO, operated by AURA, Inc., for NSF},
G. Testor $^3$
\thanks{Visiting Astronomer, European Southern Observatory, La Silla, Chile},
and S. Gim\'enez Ben\'{\i}tez $^1$
\thanks { Fellow of CIC--BA, Argentina;
Visiting Astronomer, CASLEO, San Juan, Argentina}
\\
$^1$ Facultad de Cs. Astron\'omicas y Geof\'{\i}sicas,
Universidad Nacional de La Plata, Paseo del Bosque s/n, 1900 La Plata, Argentina;\\ virpi@fcaglp.unlp.edu.ar; sixto@fcaglp.unlp.edu.ar\\
$^2$ Lowell Observatory, 1400 West Mars Hill Road, Flagstaff, AZ 86001;
massey@lowell.edu\\
$^3$ Observatoire de Paris, section de Meudon, F-92195 Meudon Cedex, DAEC,
France; Gerard.Testor@obspm.fr
}

\maketitle

\begin {abstract}
We present a study of optical spectra of the Wolf--Rayet star 
AzV 336a (= SMC WR7) in the Small Magellanic Cloud.
Our study is based on data obtained at several Observatories between 1988 
and 2001. We find SMC WR7 to be a double lined WN+O6 spectroscopic binary 
with an orbital period of 19.56 days.
The radial velocities of the He absorption lines of the O6 component
 and the strong He{\sc ii} emission at $\lambda$4686\AA\ of the WN component 
describe antiphased orbital motions. However, they show a small phase shift 
of $\sim$ 1 day.  We discuss possible explanations for this phase shift.
The amplitude of the radial velocity variations of He {\sc ii} emission is
twice that of the absorption lines. The binary components have fairly high
minimum masses, $\sim$ 18 $\modot$ and 34 $\modot$ for the WN and O6 components,
respectively. 

\end{abstract}


\section {Introduction}
Many of the most luminous stars in the central cluster of the 30 Doradus 
Supergiant H{\sc ii} region in the Large Magellanic Cloud are stars with 
emission lines of Nitrogen and Helium in their spectra (Massey \& Hunter 1998),
 classified as Wolf--Rayet (WR) stars of WN type. The luminosities of 
these stars in 30 Dor, when compared with numerical evolutionary tracks of 
massive stars, would correspond to stars of initial masses of 80--120 $\modot$.
However, no stars more massive than $\sim$60$\modot$ are known from studies of
binary star orbits, the most massive at present being R136-038, an eclipsing
O3 star in the R 136 cluster (Massey, Penny \& Vucovich 2001).

  In our Galaxy, the most massive star known in a binary system, with a mass
of 50-60 $\modot$, is the WN type star HD 92740 in the Giant Carina H{\sc ii} 
region (cf. Schweickhardt et al. 1999, and references therein). This seems to
suggest that WN stars, at least those associated with Giant H{\sc ii} regions,
are related with the upper limit of the stellar masses. Indeed, the new
stellar evolutionary tracks taking into account rotation (cf. Meynet \& Maeder
2000), predict high mass loss rates at young age for the most massive stars.
Empirical determination of stellar masses from binary star orbits
are  needed for stars with WN spectra in H{\sc ii} regions to elucidate the
role played by these stars at the upper mass limit.

In this paper we present a radial velocity study of a star with WR spectrum
in an H{\sc ii} region in the Small Magellanic Cloud (SMC), namely SMC WR7,
showing it to be a double-lined spectroscopic binary with components of 
high minimum masses.
 
In their search for stars with WR spectra in the SMC, Azzopardi \& Breysacher 
(1979)  found a star that showed broad He{\sc ii} 4686 emission in the 
optical spectrum. The star was called SMC/AB7, and interpolated with the 
number 336a in the catalog of SMC members (Azzopardi \& Vigneau 1979).
In their recent new survey for WR stars in the SMC, Massey \&
Duffy (2001) proposed to use the denomination of SMC WR7 instead, 
according to the IAU nomenclature recommendations, which we will follow here.

\begin{table*}
\caption{Observational details used for digital (CCD) spectra of SMC WR7}
\begin{center}
\begin{tabular}{cllllcccc}
\hline
Nr.&Observatory&Epoch(s)& Telescope& Spectrograph& disp.& $\Delta\lambda$& exp.time& S/N\\
   &       &          &  &             & (\AA/px)& (\AA)&  (min)& \\
\hline
1& CTIO  &1988 Nov.& 1-m   & 2DF   &  .4  &      3750-5050 &   30 &  20\\
2& CTIO  &1990 Dec.& 1-m   & 2DF   &  .4  &      3750-5050 &   60 &  35\\
3& ESO   &1992 Dec.& 1.5-m & B\&C  &   1.9  &      3700-7100 &   20 & 80 \\
4& CASLEO &1994 Jan.& 2.1-m & REOSC &  2.2  &       3800-6000&    15&   80\\
5& ESO    &1995 Jan.&  1.5-m &  B\&C  & .5  &        3900-4900 &   30& 30 \\
6& ESO    &1995 Dec.&  1.5-m &  B\&C  & .5  &       3950-4950 &   60& 40\\
7& CTIO   &1996 Oct.&  1.5-m &  Cass. & 1.1  &       3750-5050 &   10& 45\\
8& CASLEO &1996 Dec., 1997 Dec.& 2.1-m &B\&C& 2.2 &   3900-5000 &   30&  120\\
9& CASLEO &1996 Jan., 1998 Sep.&2.1-m & REOSC &.3 &   4620-4750 &   45& 30 \\
10& CTIO &1999 Jan.&   4.0-m &  R-C   &  .4  &     3700-5000  &  10& 100\\
11 & CTIO &1999 Oct., 2000 Oct.&1.5-m &Cass. &.6&   4100-4750 &   15& 45\\
12 &CASLEO &1999 Nov., 2000 July, Sep.&2.1-m & REOSC &1.8 & 3900-5500 &60&200\\
\hline
\end{tabular}
\end{center}
\label{tab01}
\end{table*}

 Azzopardi \& Breysacher (1979) considered the spectrum of SMC WR7 to be of 
peculiar WN3 type, because no N emission lines were observed. The presence 
of a companion was inferred 
from the observed strong continuum in the spectrum. Absorption lines 
were subsequently detected in the spectrum of SMC WR7 by Moffat (1988) and 
Conti et al.  (1989). Moffat (1988) assigned an approximate spectral type O7: 
for the absorption line spectrum, and also found the radial velocity of the 
absorption and emission lines to be variable, but could not determine a 
binary period. Massey \& Duffy (2001) classify the emission line spectrum 
as WN2, and the absorption line spectrum as O6 type.

SMC WR7 lies embedded in the bright H{\sc ii} region
N76-A (Henize 1956), and is one of the few Pop.I stars surrounded 
by extended nebular emission of He{\sc ii} 4686 (cf. Testor \& Pakull (1989);
 Niemela, Heathcote \& Weller 1991). The high ionization of the nebula led 
Pakull \& Bianchi (1991) to propose a very high effective temperature for 
the WN star in SMC WR7.

Massey et al. (2000) studied the surrounding OB association Hodge~53 
predicting a very high progenitor mass ($>50$) for SMC~WR7 based upon 
the turn--off mass of the cluster. They also determined an absolute magnitude 
$M_V= -5.9$, and a lower limit for the bolometric correction (B.C.) of 
$\sim -4.5$.  However, the coevality of starformation within Hodge~53  
was ranked $'$questionable$'$ by Massey et al. (2000), and the binary 
nature of SMC~WR7 was not yet fully recognized.

\section {Observations}
We have obtained 69 digital optical spectral CCD images of SMC WR7, mainly 
in the blue spectral region,
with several telescopes and spectrographs between 1988 and 2001 at Cerro Tololo
Interamerican Observatory (CTIO) and European Southern Observatory (ESO) in
Chile, and the Complejo Astron\'omico El Leoncito (CASLEO\footnote{
 CASLEO is operated under agreement between CONICET, SECYT, and the
    National Universities of La Plata, C\'ordoba and San Juan, Argentina.}
) in Argentina. 
The telescopes and instrumental configurations are listed in Table~\ref{tab01}. Our main
aim was to determine the radial velocity orbit of this WN+O binary.

\begin{figure*}
\vspace {17cm}
\includegraphics{rv2r.ps}
\caption {Radial velocity variations of He{\sc ii} 4686
emission (filled circles) of the WN component, and of the He absorptions 
of the O6 component (open circles) in the SMC WR7 binary system,
phased with the period of 19.56 days. Note the phase shift of the radial
velocity curves, which have a common origin in HJD 2447468.0}
\label{fig01}
\end{figure*}

One-dimensional spectra were extracted from our two-dimensional spectral images 
using IRAF (CTIO and CASLEO spectra) or MIDAS (ESO spectra) routines. 
These spectra were subsequently wavelength 
calibrated for the determination of positions of spectral lines. 
Radial velocities for all spectra were determined using IRAF routines. 
For the absorption lines we fitted gaussian profiles to the lines. Because 
the nebular emission is strong in hydrogen Balmer absorptions, we chose to use 
only He lines, mainly He{\sc ii} absorptions, for the mean radial velocities of
the O type component. Depending on the observed wavelength range, the radial
velocity of the O type component was determined as an average of
the lines of He{\sc ii}$\lambda\lambda$ 4200,\,4541,\,5411, occasionally
including He{\sc i}$\lambda\lambda$ 4026,\,4387,\,4471,\,5875 \AA\ .

The spectrum of the WN component is dominated by the 
strong emission of He{\sc ii} 4686 \AA\ , the only emission line for which
we could determine radial velocity values in all of our spectra. The radial 
velocities for this emission were determined both by fitting a gaussian, 
and determining the line center. When these two values were sensibly different,
which happened when the emission appeared asymmetrical, then the line center 
was preferred, otherwise a mean of the two was used.
The mean radial velocities of the absorption lines and the He{\sc ii} 4686 
emission are listed in Table~\ref{tab02}.

\section {Results and their discussion}

Early results of part of our observations showed that SMC WR7 indeed is a 
binary with a probable mass ratio of $\sim$ 0.5 (Niemela 1994); 
and a preliminary orbital solution (Niemela \& Morrell 1999) indicated very
massive binary components.

\subsection {The period}
Clearly the data in Table~\ref{tab02} confirm the variability of the radial velocities.
We have searched for periodicities in the radial velocity variations of
both the absorptions and of the He{\sc ii} emission listed in Table~\ref{tab02}. 
We used the algorithm published by Cincotta et al.  (1995). For the 
radial velocity variations of the He{\sc ii} emission we also
included the velocities published by Moffat (1988). We find the best period 
for the radial velocity variations of the absorption lines to be 
P=19.563$\pm$0.003 days, and that of the He{\sc ii} emission 
P=19.5600$\pm$0.0003 days. 
This latter period
appears more accurate due to the longer time baseline in adding the previously
published data. The previously published absorption line velocities 
appeared too noisy for an improvement in the period. Thus we have adopted 
the orbital period of SMC WR7 to be 19.560 days. 

\subsection {The radial velocity orbit}
The radial velocities of the absorptions and the He{\sc ii} emission listed
in Table~\ref{tab02} describe opposite orbital motions when phased with the period of
19.56 days, thus confirming that SMC WR7 is a double lined O+WN binary system. 
However, there appears a small phase lag of $\sim$ 1 day between the two
radial velocity curves. This is illustrated in Figure~\ref{fig01}, which depicts the
radial velocities phased with the period of 19.56 days adopting a common origin
for the phases. In this figure the maxima and minima of the radial velocities of
the He{\sc ii} emission and the absorption lines do not coincide exactly as
expected from opposite orbital motion.

Phase lags between the He{\sc ii} 4686 emission line and the orbit defined by
the absorption lines of the binary companion are observed in other WR+OB binary
systems,  e.g. WR97 in our Galaxy (Niemela, Cabanne \& Bassino 1995). However,
the origin of these phase lags is not understood.
Also in active binaries with compact components, e.g. the cataclysmic binaries 
and X-ray binaries, the He{\sc ii} 4686 emission line orbit is slightly out 
of phase from the absorption line orbit. This effect is then ascribed to a 
hot spot in an accretion disc. Stars with Wolf-Rayet spectra are usually not 
thought to have discs, but (spherically symmetric?) strong winds. The phase lag
may be related to the colliding winds of the binary components.

\begin{table*}
\caption{ Journal of observations of SMC WR7 }
\begin{center}
\begin{tabular}{rrrcrrrc}
\hline
HJD &\multicolumn{3}{c}{Heliocentric Radial Velocity ($\kms$)}&HJD &
\multicolumn{3}{c}{Heliocentric Radial Velocity ($\kms$)}\\
\cline{2-4}\cline{6-8}
  & Nr.OD & O6 abs.(n) &  He{\sc ii}4686 em.& & 
Nr.OD & O6 abs.(n) &  He{\sc ii}4686 em.\\ 
\hline
7469.630 & 1 & 235(3) &    &      10432.628 &8 & 116(2) & 257 \\
7474.720 &1  &        & 394  &    10434.632 &8 & 161(2) & 113 \\
7475.729 &1  &       & 396 &      10435.628 &8 & 190(2) & 44 \\ 
7477.702 &1  &       & 328  &       10436.601 &8 & 214(2) & 19 \\ 
7479.702 &1  & 124(1) & 272 &       10437.625 &8 & 249(2) & -6 \\ 
7480.578 &1  &     & 69  &                &  &     &   \\  
7481.618 &1  &     & 66  &      10810.589 &8 & 275(3) & 3 \\  
         &   &     &     &       10811.575 &8 & 277(3) & 16 \\ 
8249.592 &2  & 270(5) & -51 &                &  &     &   \\  
8250.592 &2  & 272(5) & 13  &       10966.918 &8 &     & -27\\ 
8251.587 &2  & 216(5) & 148 &                &  &     &   \\  
8252.587 &2  & 168(4) & 214 &      11077.649 &9 &     & 324 \\
8253.592 &2  & 141(3) & 264 &      11078.747 &9 &     & 171 \\
         &   &     &    &          11079.724 &9 &     & 74 \\ 
8982.535 &3  & 66(6) & 355 &        11080.718 &9 &     & 81 \\ 
8982.643 &3  & 62(3) & 327 &       11083.706 &9 &     & 0 \\  
8983.552 &3  & 88(2) & 328 &        11084.689 &9 &     & -71  \\
8983.594 &3  & 85(3) & 267:&       11085.689 &9 &     & -75 \\  
8983.619 &3  & 80(2) & 287  &                 &  &     &    \\ 
         &   &    &     &         11182.543 &10 & 272(4) & -10 \\
9372.602 &4  &    & 347 &        11183.527 &10 & 278(4) & 36 \\
9373.546 &4  &    & 341 &        11185.527 &10 & 262(4) & 139 \\
9374.558 &4  &    & 341  &                &   &     &   \\ 
         &   &    &      &       11470.527 &11 & 126(3) & 136 \\
9742.542 &5  & 106(2) & 322 &      11471.660 &11 & 192(3) & 122 \\
9743.533 &5  & 103(3) & 382 &      11472.493 &11 & 196(1) & 67 \\
9744.531 &5  & 93(3)  & 383 &      11474.614 &11 & 240(1) & -27 \\
9744.555 &5  & 78(2)  & 386 &      11475.705 &11 & 288(3) & -10 \\
9745.528 &5  & 72(2)  & 377 &      11477.563 &11 & 295(3) & 67 \\
9745.551 &5  & 105(3) & 389 &      11479.497 &11 & 239(3) & 154 \\
         &   &     &     &         11480.515 &11 & 209(3) & 213 \\
10077.537 &6  & 72(3) & 387 &                &   &     &    \\
10077.563 &6  & 92(2) & 392 &      11496.620 &12 & 258:(3)& 71 \\
10079.541 &6  & 107(3) & 308 &               &   &     &   \\ 
10080.544 &6  & 104(3) & 232 &     11751.805 &12 & 287(3) & 86 \\
10081.591 &6  & 142(2) & 180 &     11753.784 &12 & 217(1) & 201 \\
          &   &     &     &               &   &     &   \\ 
10086.620 &9  &     & 26  &      11806.746 &12 & 247(3) & 14 \\
          &   &     &     &                &   &     &   \\ 
10383.579 &7  & 215(2) & 72  &      11827.496 &11 & 289(3) & -19\\
10385.497 &7  & 189(3) & 173 &      11828.511 &11 & 271(2) & 0 \\ 
10386.495 &7  & 142(2) & 230 &      11829.500 &11 & 270(2) & 32 \\
10387.493 &7  &     & 261 &  & & & \\
\hline
\end{tabular}

Notes: HJD $=$ Heliocentric Julian Date\,$-\,2\,440\,000$\,d\\
Nr. OD refers to the observational details listed in Table~\ref{tab01}.\\
(n) is the number of He absorption lines included in the mean velocity of the
O6 component\\
\end{center}
\label{tab02}
\end{table*}

\begin{table}
\caption{ Preliminary Orbital Parameters for SMC WR7.}
\begin{center}
\begin{tabular}{ccc}
\hline
         & abs.&He{\sc ii}4686 em.\\
\hline
&&\\
$a \sin i$ [R$_\odot$]  & 39$\pm$1  & 75$\pm$1  \\
$K$ [km s$^{-1}$]       &  101$\pm$2 & 196$\pm$4 \\
$V_{o}$[km s$^{-1}$]    & 172$\pm$2  & 172$\pm$3 \\
$M \sin^3i$ [M$_\odot$] & 34$\pm$4   & 18$\pm$2   \\
$e$                     & 0.10$\pm$0.02& 0.07$\pm$0.02\\
$\omega$ [deg]           & 28$\pm$12  & 101$\pm$16 \\
$T_{o}$ [HJD] 2.440.000+&7468.0$\pm$0.6&7480.7$\pm$0.8\\
$P$ [days] & \multicolumn{2}{c}{19.560$\pm$0.0005}\\
\hline
\end{tabular}
\end{center}
\label{tab03}
\end{table}

\begin {figure*}
\vspace {8cm}
\includegraphics{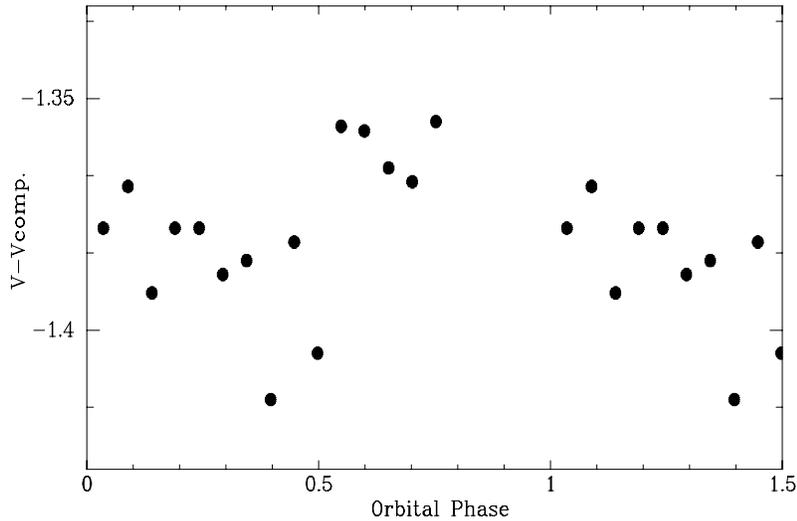}
\caption {Photoelectric light variations of SMC WR7 phased with the same 
ephemeris as the radial veolocity variations in Figure~\ref{fig01}. 
Data are from Seggewis et al. (1991).}
\label{fig02}
\end{figure*}

\begin{figure*}
\vspace {9cm}
\includegraphics{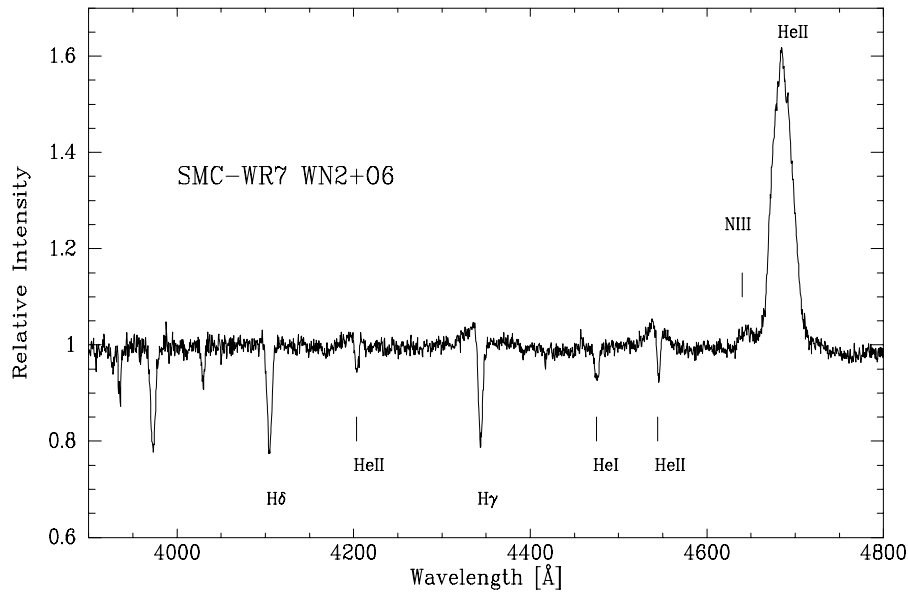}
\caption {Continuum rectified spectrum of SMC-WR7 obtained at CTIO in 1999, 
January. Absorption lines are identified below, and emission lines above the 
continuum.}
\label{fig03}
\end{figure*}

We have performed an orbital fit separately for the radial velocities of the
O star and the WN star. The orbital elements are listed in Table~\ref{tab03}.
These orbital elements are still to be considered preliminary, 
since the observed
phase lag between the absorption line orbit and the He{\sc ii} 4686 emission
casts some doubts on this last line as representative of the true orbital
motion of the WN component. We note that the minimum masses of the binary 
components appear to be quite high, 34$\modot$ and 18$\modot$ for the 
O6 and WN2 components, respectively. With such high minimum masses we would
expect to observe light variations, if not eclipses.

\begin {figure*}
\vspace {8cm}
\includegraphics{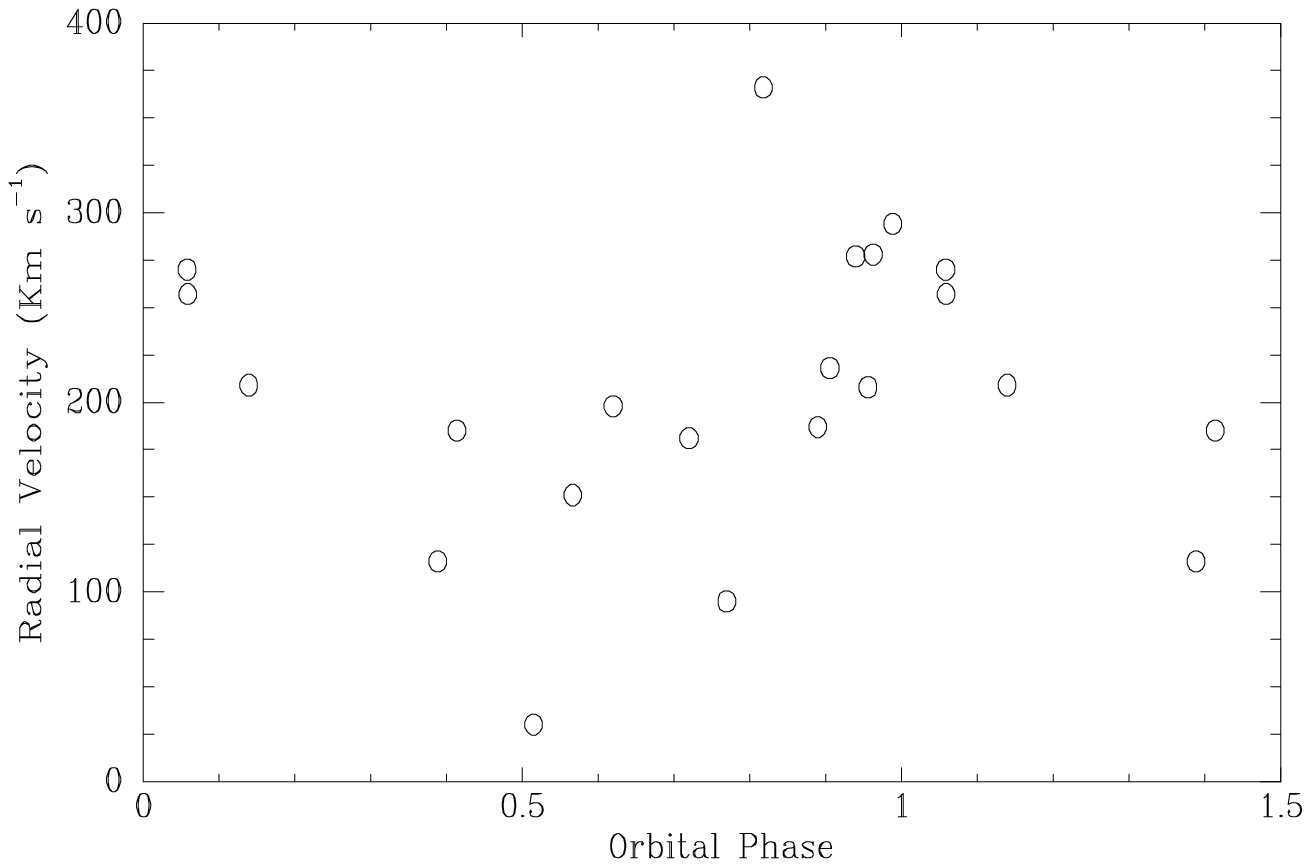}
\caption {Radial velocity variations of the faint {\sc niii} emission in the
spectrum of SMC WR7 phased with the same
ephemeris as the radial veolocity variations in Figure~\ref{fig01}. }
\label{fig04}
\end{figure*}

From photoelectric photometry Seggewiss
et al. (1991) found SMC WR7 to be slightly variable. In Figure~\ref{fig02} we have 
plotted these light variations with the same ephemeris as the radial velocity 
curves in Figure~\ref{fig01}. The minimum light then occurs just after 
the WR star passes in front of the system.
This could be a wind eclipse, but
more numerous data are needed to confirm the nature of the light variations. 
A wind eclipse would indicate an orbital inclination of at most $\sim60\deg$, 
which would bring the individual masses of the binary components to $28\modot$
for the WN component, and $54\modot$ for the O6 component.

SMC WR7 has also been observed 
by the Optical Gravitational Lensing Experiment (OGLE) (cf. Udalski et al. 1998)
where it appears as the star SMC$\_$SC9 37124. OGLE did not detect photometric
variations of SMC WR7 in their 14 B, 23 V, and 108 I broad band observations
to within 0.015, 0.019 and 0.024 mag in each band, respectively.
However, since the individual data are not published, their distribution
according to the binary orbit is not known.

\subsection {Spectra of the binary components}

The blue spectrum of SMC WR7 is illustrated in Figure~\ref{fig03}, which shows a spectrum
obtained at CTIO in 1999, January. The spectral type corresponding to the 
OB absorption lines in Figure~\ref{fig03} confirms the classification as O6 from the
relative intensities of He{\sc i} 4471 and He{\sc ii} 4542 absorptions. 
The luminosity class is difficult to ascertain, since the WN emission 
dominates the He{\sc ii} 4686, which is the main luminosity indicator for early
O type spectra in the blue spectral region.
We also note that the O6 spectrum seems to dominate 
the continuum, hence the absolute magnitude $M_V$=-5.9 of SMC WR7
(Massey et al. 2000) mainly corresponds to the O6 component of the binary.

In several of our spectra there 
appears a faint emission line at $\sim \lambda $ 4640 \AA, which we identify as
N{\sc iii}. We have been able to determine the radial velocity of this feature
in 17 of our spectra. When we phased these velocities with the same ephemeris
as those in Figure~\ref{fig01}, it is clear that the N{\sc iii} emission
follows the same orbital motion as the O6 component of the binary. 
Figure~\ref{fig04} illustrates the radial velocity variations of the 
N{\sc iii} emission in the spectrum of SMC WR7.
Given the high absolute magnitude, the most probable spectral classification 
of the absorption line component then is O6I(f). 

Otherwise, the emission line spectrum shows only lines of He{\sc ii}. Thus we
keep the WN2 classification for the emission line spectrum (cf. Massey \& 
Duffy 2001).

\subsection {Comparison with the theoretical WNE mass--luminosity relation.}

If stars with Wolf--Rayet spectra are bare He--burning cores, they should obey
a tight mass--luminosity relation (e.g. Schaerer \& Maeder 1992). In this
relation, the high minimum mass of the WN2 component of SMC~WR7, namely 
$18 M_\odot$, would imply M$_{bol}$ higher than -9.5. 

Considering that the WN2 star of the SMC WR7 binary appears as the source 
of the very high ionization in the H{\sc ii} region N76-A, which shows
strong nebular He{\sc ii}$\lambda$4686~\AA\ emission, 
Pakull \& Bianchi (1991) estimated a black body Zanstra temperature of 80kK
for the WN2 star.
Such a high temperature implies a large B.C., certainly
higher than the minimum B.C. $\sim\,-4.5$ determined by Massey et al. (2000).
Adopting the approximate relation 
between B.C. and temperature published by Vacca et al.(1996), results in
B.C.= -5.8 for the WN2 star. This is in keeping with the average 
B.C.$\sim$\,-6.0 for 
WNE stars found previously (cf. Massey et al. 2000, and references therein).

The OGLE photometry of SMC~WR7 gives V=~13.221 and B-V=~-0.194. Because the
O6 component dominates the visual light, the intrinsic (B-V)$_o$=\,-0.32.
Adopting the distance modulus 18.9 for the Small Magellanic Cloud
(e.g. van den Bergh 2000), then results in M$_v$= -6.1 for the binary system. 
This is similar to the previously published values  M$_v$= -5.9 (Massey et al.
2000), and M$_v$= -6.2 found by Crowther (2000), who also
estimated M$_v$= -5.2 for the WN component of the binary. If this component
contributes only $30\%$ to the optical light of the system (cf. Pakull \& 
Bianchi 1991), then the WN2 star has M$_{v}\sim-4.6$. With the B.C.= -5.8
(see above), this star would then have M$_{bol}$= -10.4. 
Within the uncertainties, this value corresponds for a star of $28\modot$
according the mass--luminosity relation for models of WNE stars 
(Scharer \& Maeder 1992), indicating an orbital inclination close to
$\sim$60~degrees for the SMC WR7 binary system. Further discussion
on the mass--luminosity relation shall await a careful photometric
analysis of the SMC~WR7 binary system in order to establish a reliable 
estimate of the orbital inclination. 

\section {Summary}

From spectral observations of SMC WR7 over several years, we find the
following:
\begin{enumerate}
\item Opposite radial velocity variations of the absorption lines and 
He{\sc ii}~$\lambda$4686 \AA $\,$ emission show this star to be a double lined 
O6+WN2 spectroscopic binary system.
\item The most probable period of the radial velocity variations is 19.560 days.
\item In this period, the radial velocity orbit of 
He{\sc ii}~$\lambda$4686~\AA 
$\,$ emission describes an orbit with a small phase lag of $\sim$ 1~day
relative to the orbit defined by the absorption lines.
\item Minimum masses of the binary components are quite high, 18M$\odot$ and
34M$\odot$ for the WN and O6 components, respectively.
\item Published photoelectric data of SMC WR7 phased with the 19.560 days
period, may indicate a wind eclipse of the O6 star when the WN component 
is in front of the system, precluding high orbital inclinations.
\item If the WN2 component obeys the theoretical mass--luminosity relation
for WNE stars (Schaerer \& Maeder 1992), an orbital inclination of the order
of 60 degrees is predicted.
\end{enumerate}

\section{acknowledgements}
We thank the directors and staff of CASLEO, CTIO and ESO for the use of their
facilities over many successful observing runs.
We also thank Nidia Morrell for a spectrogram of SMC WR7.
Interesting comments by an anonymous referee contributed to improve 
the discussion.
The use at CASLEO of the CCD and data acquisition system
supported under U.S. NSF grant AST-90-15827 to R. M. Rich is acknowledged.

\end{document}